# Fermi surface nesting and magnetic structure of ErGa$_3$

M. Biasini, G. Ferro

*ENEA, Via don Fiammelli 2 40129 Bologna, Italy*

G. Kontrym-Sznajd and A. Czopnik

*W. Trzebiatowski Institute of Low Temperature and Structure Research,*

*P.O.Box 937 Wroclaw, Poland.*

**More details in Phys. Rev. B (2002)**

A three dimensional mapping of the Fermi Surface (FS) of the rare-earth compound ErGa$_3$ was determined via measurements of the angular correlation of the electron-positron annihilation radiation. The topology of the electronlike FS does show nesting properties which are consistent with the modulated antiferromagnetic structure of the system.

We determine the density of states at the Fermi energy N(E$_F$) and the electronic contribution to the gamma constant to be N(E$_F$) =16 states/Ryd/cell and $\gamma$=2.7 (mJ/mole K$^2$), respectively.

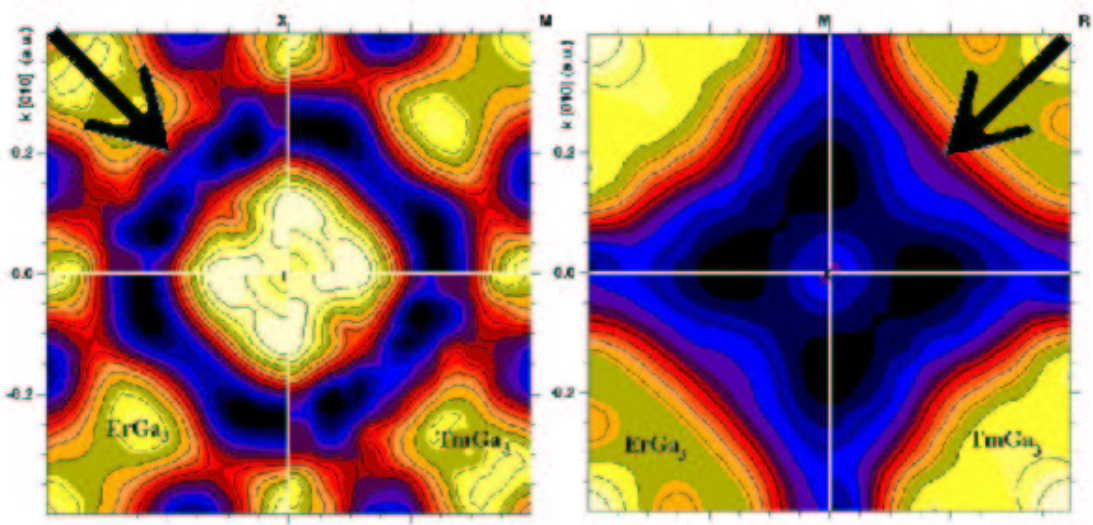

Densities $\rho(\mathbf{k})$ of ErGa$_3$ and TmGa$_3$ in the k$_z$=0 and k$_z$= $\pi$/a plane of the Brillouin zone reconstructed from 2D ACAR data. . The arrow highlights the nesting feature attributed to TmGa$_3$ [1] and ErGa$_3$ (left and right sides, respectively).